\documentclass[pra,twocolumn,showpacs]{revtex4}
\usepackage{graphicx}
\def\Journal#1#2#3#4{{#1} {\bf #2}, #3 (#4)}
\def\PR{\em Phys. Rev.}
\def\PRL{\em Phys. Rev. Lett.}
\def\PRA{{\em Phys. Rev.} A}
\def\JMP{\em J. Math. Phys.}

\def\EPJD{{\em Eur. Phys. J.} D}

\newcommand{\n}{\nonumber}
\newcommand{\bn}{\begin{eqnarray}}
\newcommand{\en}{\end{eqnarray}}
\newcommand{\h}{\hspace}
\begin{document}
\title {Interference of a Tonks-Girardeau Gas on a Ring}
\author{Kunal K. Das}
\email{kdas@optics.arizona.edu}
\author{G. John Lapeyre}
\email{Lapeyre@physics.arizona.edu}
\author{Ewan M. Wright}
\email{Ewan.Wright@optics.arizona.edu}
 \affiliation{Optical
Sciences Center and Department of Physics, University of Arizona,
Tucson, AZ 85721}
\date{\today}
\begin{abstract}
We study the quantum dynamics of a one-dimensional gas of
impenetrable bosons on a ring, and investigate the interference
that results when an initially trapped gas localized on one side
of the ring is released, split via an optical-dipole grating, and
recombined on the other side of the ring. Large visibility
interference fringes arise when the wavevector of the optical
dipole grating is larger than the effective Fermi wavevector of
the initial gas.
\end{abstract}
 \maketitle

\section{Introduction}

Recently, theoretical arguments have been presented
\cite{MooreMeystre,KetterleInouye} to demonstrate that several
stimulated processes for matter waves such as four-wave mixing,
superradiance and matter-wave amplification can be achieved in
degenerate fermion gases as well as in Bose-condensed gases.
Although such collective phenomena are often interpreted as
characteristic of Bose-condensed systems, these theoretical
studies show that this need not be the case. One of the earliest
examples of matter-wave coherence was the demonstration of
interference fringes in the density profile produced by colliding
two Bose condensates (BECs) \cite{Andrews}. While it is true that
the existence of off-diagonal long range order (ODLRO) in a system
automatically implies first-order coherence, and density fringes
may thus arise in a suitably designed interference experiment,
this does not imply that if density fringes are observed the
system under consideration automatically possesses ODLRO. In fact,
such interference fringes can also arise for suitably prepared
fermionic or thermal atom sources.

Our goal in this paper is to show that strong interference fringes
can arise in a one-dimensional (1D) gas of impenetrable bosons, a
Tonks-Girardeau (TG) gas, the conditions for which are essentially
opposite from those required for BEC \cite{Ols98,PetShlWal00},
namely, the regime of low temperatures and densities and large
positive scattering lengths where the transverse mode becomes
frozen and the many-body Schr\"{o}dinger dynamics becomes exactly
soluble via a generalized Fermi-Bose mapping theorem
\cite{Gir60,Gir65}. Even at zero temperature the TG gas does not
display BEC into a single orbital, the condensate fraction varying
as $f\approx 1/\sqrt{N}$ for a system of $N$ particles
\cite{Lenard,GirWriTri01}. The TG gas is therefore a good model
system in which to study interference effects in non-condensed
gases in 1D, which are currently of relevance due to experimental
efforts to fabricate atomic waveguides for matter wave
interferometers \cite{HinBosHug98,ThyWesPre99}. Due to the
intimate relation between the TG gas and a gas of free fermions
via the Fermi-Bose mapping our results also apply to the latter.

A simple model to study the dynamics of a TG gas is in a ring
geometry which automatically imposes periodic boundary conditions.
For the study of interference in 1D this geometry has the
advantage that atomic wavepackets can be split and recombined
without losing the strictly 1D nature of the system. In
particular, we investigate the interference that results when an
initially trapped gas localized on one side of the ring is
released, split via an optical-dipole grating, and recombined on
the other side of the ring. We study the dependence of the
resulting interference fringes on various parameters such as the
wavevector of the optical dipole grating and the number of atoms
in the wavepacket.

The remainder of this paper is organized as follows: In Section
\ref{sec:basicmodel} we set up our basic model, describe the
Fermi-Bose mapping crucial for the description of a TG gas, and
consider various conditions that need to be satisfied by our model
for a consistent physical description. In section
\ref{sec:interference} we present numerical and analytic results
for interference patterns and discuss experimental feasibility.
Finally, we conclude with a discussion of the implications of our
results and potential applications.
\section{Basic model for interference}\label{sec:basicmodel}
In this section we describe our basic model of a TG gas on a ring,
and describe our method of solving the quantum dynamics of the
problem using the Fermi-Bose mapping.
\subsection{Tonks-Girardeau gas on a ring}
The fundamental model we consider comprises of a 1D gas of $N$
hard core bosonic atoms on a ring. This situation may be realized
physically using a toroidal trap of high aspect ratio
$R=2L/\ell_{0}$ where $2L$ is the toroid circumference and
$\ell_{0}$ the transverse oscillator length
$\ell_{0}=\sqrt{\hbar/m\omega_0}$ with $\omega_0$ the frequency of
transverse oscillations, assumed to be harmonic. The transverse
trap potential is assumed to be radially symmetric about an axis
consisting of a circle  on which the trap potential is minimum.
The longitudinal (circumferential) motion can be described by a 1D
coordinate $x$ along the ring with periodic boundary conditions
applied. Then at zero temperature the quantum dynamics of the
system is described by the time-dependent many-body
Schr\"{o}dinger equation (TDMBSE) $i\hbar\partial\Psi_B/\partial
t=\hat H\Psi_B$ with Hamiltonian
\begin{equation}\label{eq1}
\hat{H}=-\frac{\hbar^2}{2m}\sum_{j=1}^{N}\frac{\partial^2}
{\partial x_{j}^{2}} +V(x_{1},\cdots,x_{N};t)  .
\end{equation}
Here $x_j$ is the 1D position of the $j^{\it th}$ particle,
$\psi_B(x_{1},\cdots,x_{N};t)$ is the N-particle wave function
with periodic boundary conditions
\begin{equation}
\Psi_B(x_1,x_2\ldots ,x_j+2L,\ldots ,x_N) = \Psi_B(x_1,x_2\ldots
,x_j\ldots ,x_N ) ,
\end{equation}
which is also symmetric under exchange of any two particle
coordinates in keeping with the Bose nature of the atoms, and the
many-body potential $V$ is symmetric (invariant) under
permutations of the particles. The two-particle interaction
potential is assumed to contain a hard core of 1D diameter $a$.
This is conveniently treated as a constraint on allowed wave
functions $\Psi_B(x_{1},\cdots,x_{N};t)$:
\begin{equation}\label{eq2}
\Psi_B=0\quad\text{if}\quad |x_{j}-x_{k}|<a\quad,\quad 1\le j<k\le
N  ,
\end{equation}
rather than as an infinite contribution to $V$, which then
consists of all other (finite) interactions and external
potentials. Here we explicitly consider the case $a\rightarrow 0$
corresponding to a gas of impenetrable point bosons.
\subsection{Fermi-Bose mapping}
To construct time-dependent many-boson solutions of Eq.
(\ref{eq1}) we employ the Fermi-Bose mapping
\cite{Gir60,GirWri00a,RojCohBer99} and start from fermionic
solutions $\Psi_{F}(x_{1},\cdots,x_{N};t)$ on a ring of length
$2L$ of the TDMSE $i\hbar\partial\Psi_F/\partial t=\hat H\Psi_F$
which are antisymmetric under all particle pair exchanges
$x_{j}\leftrightarrow x_{k}$, hence all permutations. Next
introduce a ``unit antisymmetric function"
\begin{equation}\label{eq3}
A(x_{1},\cdots,x_{N})=\prod_{1\le j<k\le N}\text{sgn}(x_{k}-x_{j})
,
\end{equation}
where $\text{sgn}(x)$ is the algebraic sign of the coordinate
difference $x=x_{k}-x_{j}$, i.e., it is +1(-1) if $x>0$($x<0$).
For a given antisymmetric $\Psi_F$, define a bosonic wave function
$\Psi_B$ by
\begin{equation}\label{eq4}
\Psi_{B}(x_{1},\cdots,x_{N};t)=A(x_{1},\cdots,x_{N})\Psi_{F}(x_{1},\cdots,
x_{N};t) ,
\end{equation}
which defines the Fermi-Bose mapping. Then $\Psi_B$ satisfies the
hard core constraint (\ref{eq2}) if $\Psi_F$ does, is totally
symmetric (bosonic) under permutations, and obeys the same
boundary conditions \cite{Gir60,Gir65}. In the Olshanii limit
\cite{Ols98} (low density, tight confined toroidal trap, large
scattering length) the dynamics reduces to that of the
impenetrable point Bose gas, the $a\rightarrow 0$ limit of Eq.
(\ref{eq2}). Then under the assumption that the many-body
potential $V$ of Eq. (\ref{eq1}) is a sum of one-body external
potentials $V(x_{j},t)$, the solution of the fermion TDMBSE can be
written as a Slater determinant \cite{Gir60,RojCohBer99}
\begin{equation}\label{eq5}
\Psi_{F}(x_{1},\cdots,x_{N};t)=\frac{1}{\sqrt{N!}}
\det_{(n,j)=(0,1)}^{(N-1,N)}\phi_{n}(x_{j},t)
,
\end{equation}
where the $\phi_{n}$ are orthonormal solutions of the single
particle time-dependent Schr\"{o}dinger equation (TDSE)
\begin{equation}\label{eq6}
i\hbar\frac{\partial\phi_n(x,t)}{\partial t}\! =\!\! \left
[-\frac{\hbar^2}{2m}\frac{\partial^2} {\partial x^{2}}
+\!V(x,t)\right ]\phi_n(x,t).
\end{equation}
It then follows that $\Psi_F$ satisfies the TDMBSE, and it
satisfies the impenetrability constraint (vanishing when any
$x_{j}=x_{\ell}$) trivially due to antisymmetry. Then by the
mapping theorem $\Psi_B$ of Eq.~(\ref{eq4}) satisfies the same
TDMBSE. The form of the single particle wavefunctions
$\phi_{n}(x,t)$ will depend on the choice of potential $V(x,t)$.

The utility of the Fermi-Bose mapping for the present problem of a
TG gas lies in the fact that the density profiles for the Fermi
and Bose problems are identical and  are both obtained as a sum
over the modulus squared of the orbitals obtained from Eq.
(\ref{eq6})
\begin{equation}
\rho(x,t) = \sum_{n=0}^N|\phi_n(x,t)|^2. \label{rho}
\end{equation}
This result follows from the fact that $A^2(x_{1},\cdots,x_{N})=1$
and hence $|\Psi_B|^2=|\Psi_F|^2$.
\subsection{Initial condition}
\begin{figure}
\includegraphics*[width=\columnwidth,angle=0]{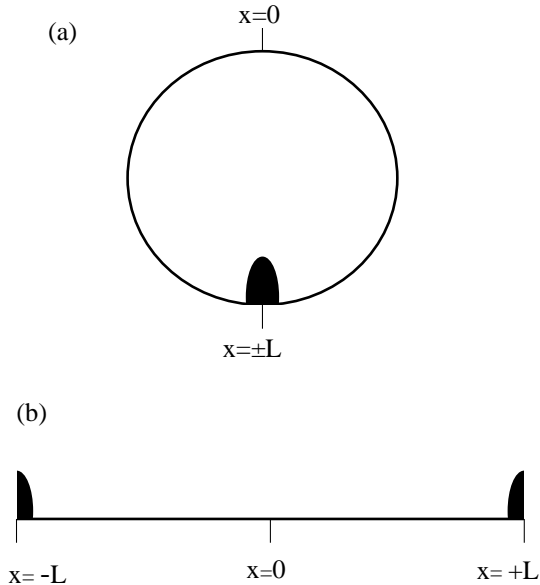}\vspace{-2cm}
\caption{(a) Our basic model consists of $N$ hard core bosons
trapped on a ring of circumference $2L$. (b) By unfolding the ring
we describe the system using a 1D coordinate $x\in [-L,L]$. The
coincident point $x=-L\equiv L$ is chosen at the center of the
initial trapped gas.} \label{figure1.RingTonks}
\end{figure}

For times $t<0$ the $N$ atoms are confined to a narrow segment of
the ring by a trapping potential $V_{HO}(x)$ which is assumed
harmonic with natural frequency $\omega$ over the spatial extent
of the initial trapped gas but drops off to zero beyond the
confines of the trapped gas (to be consistent with the periodic
boundary conditions the potential cannot be harmonic over the full
ring). The basic configuration is shown schematically in Fig.
\ref{figure1.RingTonks}(a). In order to later discuss the
time-evolution of the system, it is convenient to designate the
normal modes of an arbitrary 1D harmonic oscillator by its mean
position $\bar{x}$ and a parameter $w$ that defines the width
$x_{0}\sqrt{1+w^2}$:
 \bn u_{n}(x\!-\bar{x},w)\!=\!\frac{C_{n}}{
[1+w^{2}]^{1/4}} e^{\frac{-(x-\bar{x})^{2}}
{2(1+w^{2})x_{0}^{2}}}H_{n}\!\left[\frac{x-\bar{x}}{\sqrt{1+w^{2}}\
x_{0}}\right],\n\\C_{n}= \frac{1}
{\sqrt{\pi^{1/2}x_{0}2^{n}n!}},\label{un} \h{3.8cm}\en
 with $H_{n}$ the Hermite polynomials and
 $x_{0}=\sqrt{\hbar/m\omega}$ the single-particle ground-state width
corresponding to the initial trap potential $V_{HO}(x)$. Then the
modes of the initial trap potential are given by
$u_{n}(x\!-\bar{x},w=0)\!$, assuming that the trap is effectively
harmonic for all modes $n$ of interest. We choose our coordinates
to have the trap center at $x=L\equiv -L$. On unwrapping the ring
about $x=0$ in Fig. \ref{figure1.RingTonks}(b), the initial
Hermite-Gaussian orbitals $\phi(x,t< 0)$ are split into two parts
at the ends of the interval $[-L,L]$, and can be written as
\begin{equation}
\phi_{n}(x,t\leq 0)=(-1)^{n}[u(x+L,0)+u(x-L,0)],
\end{equation}
the factor of $(-1)^{n}$ arising from the parity-reversal
introduced by the unwrapping.  The $N$-particle fermionic ground
state $\Psi_F^{0}$ has one particle in each of the $N$ lowest
levels \cite{KolNewStr00}, and the ground-state of the TG gas is
given simply by $\Psi_B^0=|\Psi_F^0|$ \cite{GirWriTri01}, so the
Fermi and TG gases have the same ground-state density profile. The
width of the density profile of the $N$-particle ground state is $
x_N\approx\sqrt{2N}\ x_{0}$. We want this initial state to be well
localized on one side of the ring, which requires that $L\gg x_N$
or
\begin{equation}
N\ll \frac{1}{2}\left(\frac{L}{x_{0}}\right)^{2} . \label{cond1}
\end{equation}
We also define an effective Fermi wavevector $k_F$ for the initial
state by examining the highest occupied mode $n=N$. For $N\gg 1$
the asymptotic form of the Hermite polynomials yields
$\phi_N(x,0)\propto \cos(\sqrt{2N}x)$ \cite{GradRyz}, which
corresponds to a highest occupied wavevector
\begin{equation}
k_F =\frac{ \sqrt{2N}}{x_{0}} .
\end{equation}
If the initial trapped gas is released at $t=0$ and left to expand
freely it takes a time
\begin{equation}
t_{wrap} = \frac{L}{(\hbar k_F/m)} = \frac{1}{\omega\sqrt{2N}}
\left ( \frac{L}{x_{0}} \right )  , \label{twrap}
\end{equation}
for the highest excited orbital to start to wrap around the ring
of length $L$.  Equation (\ref{twrap}) shows the explicit relation
between the Fermi wavevector and the wrap time; the larger the
atom number, and hence the Fermi wavevector, the shorter the wrap
time.

Another time scale of physical significance is the Poncaire
recurrence time $t_{pr}=m(2L)^2/\pi\hbar=4(L/x_0)^2/\pi\omega$,
which is the recurrence time for the density profile and all other
physical quantities of the TG gas \cite{RojCohBer99,GirWri00a}.
Since we consider the limit $L/x_0\gg 1$ for a gas well localized
on one side of the ring $\omega t_{pr} \gg 1$, and for this paper
we consider times $t\ll t_{pr}$ thereby prohibiting any
recurrences.

A few conditions need to be satisfied for the TG gas to be
realized on a ring. For the system to be frozen in a single
transverse mode we require $N\hbar\omega \ll \hbar\omega_0$, with
$\omega_0$ the transverse oscillation frequency, or
\begin{equation}
N\ll\left (\frac{x_{0}}{\ell_0}\right )^2 , \label{cond2}
\end{equation}
$\ell_0$ being the transverse ground-state width. Furthermore, for
the initial gas to be accurately described as an impenetrable gas
of bosons we require $k_F|a_{1D}|\ll 1$, where $a_{1D}$ is the
effective 1D scattering length \cite{Ols98}. From this condition
we find
\begin{equation}
N\ll\frac{1}{2}\left(\frac{x_{0}}{a_{1D}}\right)^{2} \simeq
\frac{1}{2}\left(\frac{x_{0}a}{l_{0}^{2}}\right)^{2}.
\label{cond3}
\end{equation}
All three conditions in Eqs. (\ref{cond1}), (\ref{cond2}) and
(\ref{cond3}) need to be satisfied to realize a Tonks-Girardeau
gas initially localized on one side of the ring.
\subsection{Optical-dipole grating}
In order to produce interference from the initial trapped gas we
turn off the harmonic trap at $t=0$ and apply a temporally short
but intense spatially periodic potential of wavevector $k$. This
spatially periodic grating may be produced over the spatial extent
of the trapped gas, for example, using intersecting and
off-resonant pulsed laser beams to produce an intensity grating
whose wavevector may be tuned by varying the intersection angle,
which in turn produces a spatially periodic optical-dipole
potential for the atoms. The applied periodic potential then
produces counter-propagating scattered atomic waves, or daughter
waves, from the initial gas, or mother, with momenta $\pm\hbar k$
and these recombine on the opposite side of the ring at a time
\begin{equation}
t_{r} = \frac{L}{(\hbar k/m)}=\left(\frac{k_F}{k} \right )t_{wrap}
. \label{trec}
\end{equation}
Clearly, if we want the scattered atoms to recombine before the
initial gas wraps around the ring, $t_{r}<t_{wrap}$, we require
$k>k_F$, otherwise interference of the scattered waves will be
obscured by the wrapping.

The action of the optical dipole potential is best incorporated in
our model before unwrapping with $L\equiv -L$, by writing the
single-particle potential as a sum of the initial harmonic trap
(HO) potential and the optical-dipole potential
\begin{equation}
V(x,t) = \theta(-t)V_{HO}(x)-\delta(t)\eta\cos[k(x-L)]\
e^{-\frac{(x-L)^2}{w^2}} ,
\end{equation}
where the Heaviside function $\theta(-t)$ ensures that the
harmonic trap turns off for $t>0$ $\eta$ is the strength of the
applied periodic potential, and $w$ is the spatial extent of the
grating. In particular, the periodic grating due to the dipole
potential need only extend over the spatial extent of the initial
trapped gas, $w>x_N$, but we do require many periods of the
grating $kw\gg 1$. For simplicity in presentation here we set
$w\rightarrow\infty$. Furthermore, following Rojo {\it et. al.}
\cite{RojCohBer99} we have used a delta-function approximation for
the short pulse excitation of the periodic grating at $t=0$. Then
by integrating Eq. (\ref{eq6}) for each orbital over the
delta-kick, just after the pulse at $t=0^{+}$ each
Hermite-Gaussian mode is changed to
\begin{eqnarray} \label{initcond}
\phi_{n}(x,0^{+}) &=& e^{i\eta\cos[k(x-L)]}\phi_{n}(x,0) \nonumber \\
&=& \sum_{m=-\infty}^\infty
i^{m}J_{m}(\eta)e^{imk(x-L)}\phi_{n}(x,0) ,
\end{eqnarray}
where $J_{m}$ are Bessel functions \cite{GradRyz}.  The subsequent
quantum dynamics of the system is then traced by propagating each
orbital $\phi_{n}(x,0^{+})$ using Eq. (\ref{eq6}).

For the present discussion we assume the limit $|\eta|<1$, in
which case
\begin{equation}\label{phi0+}
\phi_{n}(x,0^{+})\approx \left [1+\frac{i\eta}{2}\left
(e^{ik(x-L)}+e^{-ik(x-L)}\right )\right ]\phi_{n}(x,0)  ,
\end{equation}
and the optical-dipole grating predominantly produces two
scattered waves with wavevectors $\pm k$ in addition to the
initial parent Hermite-Gaussian mode $\phi_{n}(x,0)$. In
particular, here we chose a value $\eta=1/2$ in which case $10$\%
of the mother wave gets transferred into each daughter, and less
than $1$\% is deflected into diffraction orders with $|m|\ge 2$.
Note also that the daughter waves are simply reduced amplitude
copies of the initial mode traveling to the left and right. We
wish to examine the presence or absence of interference between
the daughter waves at the opposite side of the ring at time
$t=t_{r}$ and how this depends on the wavevector $k$ of the
applied optical-dipole grating.
\begin{figure}
\includegraphics*[width=\columnwidth,angle=0]{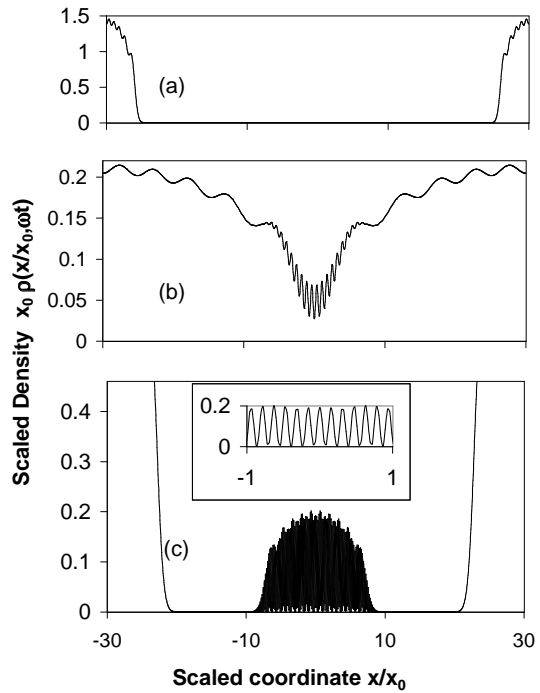}
\caption{Scaled density profiles $x_{0}\cdot\rho_B(x/x_{0},\omega
t)$ for $N=10,\eta=0.5$, $L/x_{0}=30$ and (a) the initial
condition $t=0$, (b) $x_{0}k = 0,\omega t=\omega t_{wrap}=6.7$,
and (c) $x_{0}k=20, \omega t=\omega t_{r}=1.5$, the inset shows
details of the fringes.} \label{figure2.RingTonks}
\end{figure}

\section{Interference on a ring}\label{sec:interference}
\subsection{Numerical simulations}
To set the stage for the analytic results below, Fig.
\ref{figure2.RingTonks} shows illustrative examples of
interference on a ring for $N=10$ atoms under various conditions.
To produce numerical results we solve the Schr\"odinger equation
(\ref{eq6}) for each orbital with the initial condition
(\ref{initcond}) using the split-step Fast-Fourier transform
method \cite{FleMorFei76}, and calculate the density using Eq.
\ref{rho} for the density. Figure \ref{figure2.RingTonks} shows
the scaled density profiles $x_{0}\cdot\rho_B(x/x_{0},\omega t)$
for $\eta=0.5, \ L/x_{0}=30$ and (a) the initial condition $t=0$,
(b) $x_{0}k = 0,\omega t=6.7$, and (c) $x_{0}k=20, \omega t=\omega
t_{r}=1.5$. Note that the initial density profile in Fig.
\ref{figure2.RingTonks}(a) appears split due to the choice of
origin in Fig. (\ref{figure1.RingTonks}). In Fig.
\ref{figure2.RingTonks}(b) we have allowed the initial trapped gas
to expand freely ($k=0$ is equivalent to no grating at all) and
the density is plotted at a time equal to the wrap time $\omega
t_{wrap}=6.7$. Here we see that while there is some degree of
interference the visibility is not very large, and we show in the
next subsection that these fringes tend to vanish for large
numbers of atoms. In contrast, mean-field theory, in which the
matter waves are described by a single orbital obeying a nonlinear
Schr\"odinger equation, results in a visibility close to unity.
(We discussed the failure of mean-field theory for TG gases
elsewhere in our previous paper \cite{breakdown}). By comparison
Fig. \ref{figure2.RingTonks}(c) for $x_{0}k=20, \omega t=\omega
t_{r}=1.5$ shows very clean interference fringes in the density
with wavevector $k_{int}=2k$ between the scattered waves upon
recombination at $t=t_{r}$ with almost unity visibility. As we
show below, these fringes do not vanish in the limit of large $N$
as long as $k>k_F$ is maintained.

These numerical simulations highlight our main findings: First,
for $k/k_F\ll 1$ at best low visibility interference fringes are
seen, and this clearly results from the fact that the initial
trapped gas has now wrapped around the ring at $t=t_{r}>t_{wrap}$.
Second, for $k/k_F\gg 1$ large visibility interference fringes
between the recombined scattered waves can be observed. Rojo {\it
et. al.} \cite{RojCohBer99} obtained the same criterion for the
appearance of Talbot oscillations in a 1D TG gas on a ring of size
$L$. In their case the initial state of the gas is homogeneous on
the ring with no external trap, for which $k_F=\pi N/L$, and the
Talbot oscillations in the density due to the applied grating are
undamped only for $k/k_F\gg1$.
\subsection{Analytic results}
In this section we develop an analytic approach to describe the
interference fringes for the TG gas on a ring. Immediately after
the periodic potential is applied and the harmonic trap turned
off, the single particle modes (\ref{phi0+}) for the atoms in the
unwrapped configuration assume the form
\begin{eqnarray}\label{phi0+un}
\phi_{n}(x,0^{+})\approx (-1)^{n}\left[\left
(1+\frac{i\eta}{2}e^{ik(x+L)}\right)u_{n}(x+L,0)\right.\h{7mm}\n\\\left.+
\left (1+\frac{i\eta}{2}e^{-ik(x-L)}\right)u_{n}(x-L,0)
\right].\h{2mm}
\end{eqnarray}
For each $n$ this initial condition is the same as a superposition
of waves centered at $x=\pm L$ on the infinite line $x\in
[-\infty,\infty]$ since the initial gas is well localized on one
side of the ring. The evolution of each orbital $\phi_{n}(x,t)$
may therefore be approximated as free propagation on the infinite
line for times $t<min(2t_r,2t_{wrap})$. For example, for the case
of no applied optical dipole potential $\eta=0$, at $t=t_{wrap}$
the tails of the two expanding packets start to overlap at $x=0$
corresponding to wrapping of the mother packet, whereas at
$t=2t_{wrap}$ the tails of the expanding packets start to circle
back to $x=L\equiv -L$. Therefore, for $t\ge 2t_{wrap}$ the
solution has spread over the full ring, and approximation of the
evolution of the orbitals as free expansion on the infinite line
is no longer valid.

Here we use the approximation of free expansion of the orbitals to
investigate the interference fringes at $t=t_r$ and $t=t_{wrap}$.
Then the total particle density anywhere along the ring for $t>0$
retains the simple structure shown in Eq. (\ref{rho}) i.e. it is
simply the sum of the densities due to each occupied mode. Each
initial wavefunction $\phi_{n}(x,0^{+})$ in Eq. (\ref{phi0+un})
may be separately evolved for $t>0$ using the retarded
free-particle Green's function to obtain
 \bn \phi_{n}(x,t)=\frac{1}{\sqrt{2\pi
i\omega t}} \int_{-\infty}^{\infty}dx'\ \phi_{n}(x',0^{+})
 e^{-\frac{(x-x')^{2}}{2i\omega t}}. \label{21} \en
Due to the symmetric form of the initial wavepacket the centers of
the daughter packets arrive at $x=0$ simultaneously at a time
$t_{r}$ given by Eq. (\ref{trec}).  The time-evolved functions can
be written in terms of the functions $u_{n}(x-\bar x,w)$ defined
in Eq.~(\ref{un}). At $t=t_r$ the daughter components are centered
at $x=0$ and their width increased to
$x_{0}\sqrt{1+\omega^{2}t_{r}^2}$, thus their moduli are identical
and given by $u_{n}(x,\omega t_{r})$ but their phases are
different due to opposite velocities and the fact that the
time-evolved modes acquire a spatial phase variation as they
expand. The mother packet has the same increase in width but is
still centered at $x=\pm L$ and hence its two ends have moduli
given by $u_{n}(x\pm L,\omega t_{r})$ and their phases, being
space-dependent, are different as well.

\begin{figure}
\includegraphics*[width=\columnwidth,angle=0]{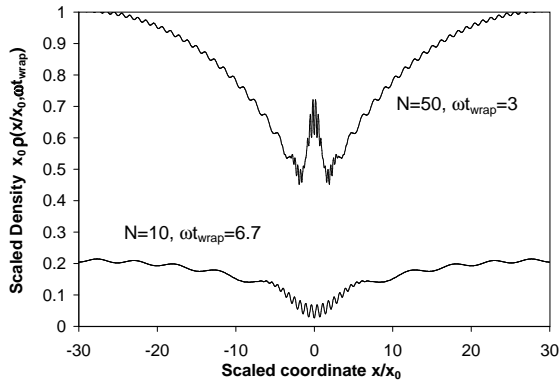}
\caption{Scaled density profiles at $t_{wrap}$ for $\eta=0$ using
the analytic expression in Eq. (\ref{term1}). It shows that for a
larger particle number $N$ there is lower
 visibility of the fringes due to the overlap of the expanding mother packet with itself .}
 \label{figure3.RingTonks}
\end{figure}

On using the time-evolved wavefunctions $\phi_{n}(x,t_{r})$ in Eq.
(\ref{rho}) we obtain an analytic expression for the density of
atoms in the neighborhood of $x=0$ which comprises three distinct
contributions
 \bn\label{rhototal}\rho(x,t_{r})=
 \rho_{m}(x,t_{r})+\rho_{d}(x,t_{r})+\rho_{md}(x,t_{r}).\en
The first term is due to the freely expanding mother packet
overlapping with itself
\bn\label{term1}
\rho_{m}(x,t_{r})=\sum_{n=0}^{N-1}\left[u_{n}^{2}(x\!\!+\!\!L,\omega
t_{r})+ u_{n}^{2}(x\!\!-\!\!L,\omega
t_{r})\right.\h{6mm}\n\\\left.+2 u_{n}(x\!\!+\!\!L,\omega
t_{r})u_{n}(x\!\!-\!\!L,\omega t_{r}) \cos(2Qx)\right].\en
with a mode-independent sinusoidal modulation of period
$\pi/Q(t_{r})= \pi x_{0}^{2}(1\!+\!\omega^{2}t_{r}^{2})/L\omega
t_{r}$ which arises form the spatial phase variation developed by
the expanding mode functions.  This would be the sole contribution
if $\eta=0$ i.e. in the absence of any daughter packets.  Figure
\ref{figure3.RingTonks} shows that the visibility of the fringes
arising from the mother packet wrapping around the ring diminishes
with increasing number of particles. The reason for this is
apparent from the above expression for $\rho_{m}$, namely, while
the first two terms representing the background being the modulus
squared of the orbitals are positive definite for all values of
$n$, for a given $n$ the sign of the product of Hermite-Gaussians
$u_{n}(x\!\!+\!\!L,\omega t_{r})u_{n}(x\!\!-\!\!L,\omega t_{r})$
in the cross term responsible for the interference fringes can
have either sign. Thus, upon summing over $n$ the value of the
cross term is in general degraded by cancellations with respect to
the background terms in Eq. (\ref{term1}). For larger number of
particles $N$ the Hermite polynomials become more oscillatory
spatially \cite{GradRyz} and we expect the cancellations to become
more complete, while the background increases like $N$, hence the
fringe contract due to wrapping of the mother packet decreases
with $N$. This physical picture of the degradation of the density
interference fringes due to destructive interference of orbitals
is in agreement with our previous discussion of the failure of
mean-field theory for a TG gas \cite{breakdown}, that is, the
quantum dynamics of a TG gas cannot be correctly captured by a
single orbital, though Kolomeisky {\it et. al.} \cite{KolNewStr00}
have shown that mean-field theory can yield the ground state
density profile.

The second term arises from the complete overlap of the daughter
packets at the recombination time $t_{r}$
\bn \label{daugh}
\rho_{d}(x,t_{r})=\frac{\eta^{2}}{2}[1+\cos(2kx)]
\sum_{n=0}^{N-1}u_{n}^{2}(x,\omega t_{r}).\en
In the case when $k\gg k_{F}$ this term dominates in the vicinity
of  $x=0$ at $t_{r}$ and produces fringes of essentially unit
visibility, and the amplitude of the modulation would increase
with particle number. This explains the strong fringes seen in
Fig. \ref{figure2.RingTonks}~(c) discussed in the previous
section. The form of the interference pattern (\ref{daugh}) is
identical to that obtained by Moore and Meystre
\cite{MooreMeystre} for a Fermi gas exposed to a Bragg grating,
and verifies within the context of an exactly soluble model that
both the TG and Fermi gases can exhibit high visibility
interference fringes.

For $k$ comparable or smaller than $k_{f}$ the clean fringes are
lost. This is mainly due to the background introduced by the
mother packet overlapping with itself. There is also a more
complicated modulation due to the overlap of the daughter packets
with the mother packet given by
\bn \rho_{md}(x,t_{r})=
 -2\eta\cos(kx)\sum_{n=0}^{N-1}u_{n}(x,\omega t_{r})\h{2.3cm}\n\\
\times\left[u_{n}(x\!\!-\!\!L,\omega t_{r})\sin(\phi\!\!+\!\!
Qx)+u_{n}(x\!\!+\!\!L,\omega t_{r})\sin(\phi\!\!-\!\!Qx)\right] .
\en
The sinusoidal modulation here has twice the period as that of
$\rho_{m}$ and in addition there is a time-dependent phase
$\phi(t_{r})=Lk/2(1\!+\!\omega^{2}t_{r}^{2})$.
 For our choice of the point of detection, diametrically opposite the center of the
initial packet, the relative importance of this contribution
diminishes with increasing particle number. This is because in the
neighborhood of $x=0$, $u_{n}(x)u_{n}(x\pm L)<u_{n}^{2}(x)$ and
this difference is enhanced as more modes are included in the sum.
Thus for a large number of particles the density distribution
around $x=0$ at time $t_r$ can be well described by the full
visibilty modulation arising from the interference of the
daughters superimposed on the background density due to each end
of the expanding initial packet.
\subsection{Experimental feasibility}
Although it is not yet possible to realize a TG gas on a ring it
is of interest to examine some parameters to assess the
possibility of experimental realization.
 Let us first consider the conditions required to achieve the impenetrable
 TG regime. Among the two constraints on the number of particles in Eq.
(\ref{cond2}) and (\ref{cond3}), the second one is more limiting
since a positive 1D scattering length requires $\ell_{0}>a$
\cite{Ols98}.  For sodium atoms with scattering length a=2.75 nm
this limits the number of particles to $N\ll 10^{-8}\times
(\nu_0^{2}/\nu)$ where $\omega_{0}=2\pi\nu_{0}$. A choice of
$\nu\sim 1$ Hz for a weak longitudinal confinement would mean that
for N=100 atoms we would still need in excess of $10^{5}$ Hz
transverse confinement frequency, which is about two orders of
magnitude above the limits of current experimental trap
frequencies. We also note that for the atoms in the daughter
packets to remain impenetrable the transverse potential should
also be tight enough to satisfy $k|a_{1D}|\ll 1$ which is relevant
for momentum kicks $k>k_{F}$ needed to observe clean fringes.

For $\nu\sim 1$ Hz the oscillator length is $x_{0}\sim 10^{-5}$~m
and hence $k_{F}\sim \sqrt{N}\times 10^{5}\ {\rm m}^{-1}$. The
wavevectors corresponding to the yellow sodium lines are about
$k\sim 10^{7}\ {\rm m}^{-1}$, thus the maximum boosts we can
realistically consider are about $k/k_{F}\sim 100/\sqrt{N}$. The
values of $k$ we used for our plots satisfy this condition.
Corresponding to this oscillator length the condition for a well
localized initial packet stated in Eq. (\ref{cond1}) will require
a ring size of $2L \sim 1$ mm, an experimentally realistic size
for atom interferometers.

At the same time, while the TG limit is very interesting it is not
a requirement for observation of interference fringes in the ring
configuration: Even if the atoms are penetrable, interference of
the scattered waves is expected anyway. If we have a dilute gas of
fermions instead of bosons we need only satisfy constraint Eq.
(\ref{cond2}) for 1D behavior which requires the much weaker
constraint $N\ll \nu_{\perp}/\nu$. Even with currently achievable
trap conditions it should be possible to realize up to $N\sim
1000$ fermionic atoms in the ring configuration and produce
interference fringes. Actual observation of the fringes however
might be difficult due to the low atom number.
\section{Summary and conclusions}
In conclusion, we have shown that a Tonks-Girardeau gas on a ring
can show large visibility interference fringes.  This demonstrates
in the framework of the exactly soluble many-body TG gas problem
that one does not need a BEC or a source possessing high-order
coherence to get interference fringes in principle. Physically,
the reason interference fringes can appear even for an incoherent
atom source is that the optical dipole grating imprints a
different spatially dependent phase on each of the diffracted
fragments and the interference fringe seen from subsequent
recombination of those fragments reflects the effect of the
imprinted phase rather than the intrinsic coherence of the initial
or mother packet. Furthermore, if the ring is rotating, the
interference fringes due to interference of the daughter waves
will shift due to the Sagnac effect: This is easily deduced by
realizing that the usual Sagnac effect will apply to each
individual orbital independent of orbital number. This implies in
turn that rotation sensors, in principle, also do not require a
coherent source of atoms.

In addition to the large visibility fringes that result from
interference of daughter waves that survive in the limit of large
$N$, fringes can also appear due to the wrapping of the mother
wave around the ring. However, the visibility of the fringes due
to wrapping was shown to decrease with increasing particle number,
thereby vanishing in the thermodynamic limit. \vspace{0.2cm}

\noindent We appreciate valuable discussions with Professor Marvin
Girardeau of the Optical Sciences Center, University of Arizona,
and Prof. Mara Prentiss of Harvard University. This work was
supported by the Office of Naval Research Contract No.
N00014-99-1-0806, and the U.S. Army Research Office.
\end{document}